\documentclass[12pt]{article}

\font\twelve=cmbx10 at 15pt
\font\ten=cmbx10 at 12pt

\renewcommand{\thefootnote}{\fnsymbol{footnote}}

\usepackage{rotating}
\usepackage{amstext}
\usepackage{amssymb}
\def\be{\begin{equation}}
\def\ee{\end{equation}}
\def\bq{\begin{eqnarray}}
\def\eq{\end{eqnarray}}
\def\beq{\begin{eqnarray}}
\def\eeq{\end{eqnarray}}
\def\ov{\overline}

\def\unal{\underline{\alpha}}

\def\gp{g^{\perp}}
\newcommand{\sect}[1]{\setcounter{equation}{0}\section{#1}}

\newcommand{\eps}{\epsilon}

\newcommand{\nm}{\nonumber}

\def\Gtilde{\tilde{G}}

\begin{document}

\begin{titlepage}

\begin{center}

\renewcommand{\thefootnote}{\fnsymbol{footnote}}

{\ten Centre de Physique Th\'eorique\footnote{Unit\'e Propre de
Recherche 7061} - CNRS - Luminy, Case 907}

{\ten F--13288 Marseille - Cedex 9}

\vspace{1.5 cm}

{\twelve LIGHT-CONE WAVE FUNCTIONS OF THE PHOTON: CLASSIFICATION UP TO TWIST FOUR}

\vspace{0.3 cm}
\setcounter{footnote}{0}
\renewcommand{\thefootnote}{\arabic{footnote}}

{\bf Gautier STOLL}

\bigskip

Fond National Suisse de la Recherche Scientifique

\vspace{1.5 cm}

{\bf Abstract}

\end{center}

The different light-cone wave functions of the photon up to twist four are defined. Some explicit expressions are extracted from results of Balitskii and al.\cite{BBK}, Ali and Braun\cite{BA}.

\vspace{\stretch{1}}

\noindent Key-Words : Non-perturbative QCD, Conformal expansion.
\bigskip

\noindent March 1999
\bigskip

\noindent CPT-99/P. 3805
\bigskip

\noindent anonymous ftp: ftp.cpt.univ-mrs.fr\\
\noindent www.cpt.univ-mrs.fr

\pagebreak

\end{titlepage}

\setcounter{footnote}{0}
\renewcommand{\thefootnote}{\arabic{footnote}}

%%%%%%%%%%%%%%%%%%%%%%%%%%%%%%%%%%%%%%%%%%%%%%%%
%%%%%%%%%%%%%%%%%%%%%%%%%%%%%%%%%%%%%%%%%%%%%%%%

\sect{Introduction}

Understanding the non-perturbative aspects of QCD is still an unsolved problem, but several ways of parameterizing these effects exist.

One method is to use the notion of "wave function" or distribution amplitude (momentum fractions of partons in particular meson state). This notion, firstly introduced by Brodsky and Lepage\cite{BL}, is specially useful when one deals with hard exclusive processes.

In an other way, the notion of wave function appears in the method of QCD light-cone sum rules\cite{B} as basic non-perturbative objects (the basic non-perturbative objects of the usual QCD sum rules are vacuum condensates \cite{SVZ}).

In the approach of Brodsky and Lepage\cite{BL}, the parton decomposition is consider in the infinite momentum frame (a mathematically equivalent approach is the light-cone quantization\cite{BPP}). In this paper, I take the point of view of Braun and al.\cite{BF, BBK, BBKT,BB, BBS}, where the wave functions are extracted from matrix elements between vacuum and a physical state (meson or photon) of a gauge invariant light-cone operator. In this approach, one obtains directly the inputs for the method of light-cone QCD sum rules; in addition, the exact equation of motion can be used.

In this paper, I classify the different wave functions of the photon up to twist-4 (the meson case has already been treated \cite{BF, BB, BBKT, BBS, GS}). Then I extract some explicit expressions from \cite{BBK, BA}.

%%%%%%%%%%%%%%%%%%%%%%%%%%%%%%%%%%%%%%%%%%%%%%%%
%%%%%%%%%%%%%%%%%%%%%%%%%%%%%%%%%%%%%%%%%%%%%%%%

\sect{General Framework}

In this paper, the photon wave functions are extracted from vacuum expectation values of operators (massless quark fields and gluon field) on the light-cone, at first order in the electric charge in an external classical electromagnetic field.

There are the two-points wave functions, extracted from this kind of matrix element:
\be
\left\langle 0 |\ov{\psi}(x)[x,-x] \Gamma \psi(-x)|0 \right\rangle_{F_{\mu\nu}} \label{2matel}
\ee
and the three points wave functions, extracted from this kind of matrix element:
\be
\left\langle 0 | \ov{\psi}(x)[x,vx] g G_{\mu\nu}(vx)[vx,-x] \Gamma \psi(-x)|0\right\rangle_{F_{\mu\nu}}
\label{3matel}
\ee
where $x$ is almost on the light cone, $v \in [0,1]$, $\Gamma$ any kind of product of $\gamma_{\mu}$ matrices and $[x,y]$ a path-ordered gauge factor along the straight line connecting $x$ and $y$ :
\be
[x,y]={\rm P}\exp \left\{ i\int^1_0 dt (x-y)_\mu [g_s A^\mu (tx+(1-t)y)
-g_e B^\mu(tx+(1-t)y)]\right\}
\ee
$A_\mu$ is the gluon field and $B_\mu$ is the electromagnetic field, which is related to the gauge-invariant classical electromagnetic tensor field 
$F_{\mu\nu}=i \exp(iqx)(\eps_\mu q_\nu-\eps_\nu q_\mu)$.

The gauge factor is a way to introduce interaction (see \cite{Bali}), it ensure the gauge invariance of these non-local matrix elements (I will omit to write it sometimes).

In order to classify these wave functions, the projector onto the direction orthogonal to $q$ and $x$ is useful:
\be 
\gp_{\mu\nu}=g_{\mu\nu} -\frac{1}{qx}(q_\mu x_\nu -q_\nu x_\mu)
\ee

The following notation will be often used:
\be
a.\equiv a_\mu z^\mu \; , \: a_* \equiv a_\mu p^\mu / (pz)
\ee

In order to keep the Lorenz-invariance and the gauge-invariance, the matrix-elements \ref{2matel} and \ref{2matel} can only be functions of
\bq
 q_\mu \nm \\
x_\mu \nm \\
F_{\mu\nu} \nm \\
\gp_{\mu\nu} \nm \\
qx \nm
\eq

$q^2$ and $x^2$ are set to zero (a non zero $x^2$ can appear explicitly for twist-4 corrections of twist-2 wave functions).

%%%%%%%%%%%%%%%%%%%%%%%%%%%%%%%%%%%%%%%%%%%%%%%%
%%%%%%%%%%%%%%%%%%%%%%%%%%%%%%%%%%%%%%%%%%%%%%%%

\sect{Definition of the different wave functions}

\subsection{Twist classification}

For local operators, the twist means the dimension minus the spin. But for non-local matrix elements, the definition of twist is a little different: it is built in analogy with the case of Deep Inelastic Scattering, where the different "twist" give contributions to different powers in the hard momentum transfer. A good description of the notion of twist can be found in \cite{JJ}. In this paper, the classification is taken in analogy to the meson wave functions (\cite{GS, BBKT, BBS}), up to twist-4.

%%%%%%%%%%%%%%%%%%%%%%%%%%%%%%%%%%%%%%%%%%%%%%%%

\subsection{Two-points wave functions}

Chiral-Even:

\bq
\left\langle 0 |\ov{\psi}(x)\gamma_\mu \psi(-x)|0 \right\rangle_{F_{\mu\nu}} 
& = &
e_\psi\int_0^1 du F_{\mu\rho}(\xi x)x^\rho C(u)
\\
\left\langle 0 |\ov{\psi}(x)\gamma_\mu \gamma_5 \psi(-x)|0 \right\rangle_{F_{\mu\nu}} &=&
e_\psi \int_0^1 du F_{\alpha\beta}(\xi x) x^\rho \eps_{\mu\nu\alpha\beta}a(u)
\eq
where $\xi=2u-1$. $x^2$ is set to zero ($x^2$ corrections are already twist 5). $e_\psi$ is the electric charge of the (massless) quarl field $\psi$.

Chiral-Odd:

\bq
\left\langle 0 |\ov{\psi}(x)\sigma_{\mu\nu}\psi(-x)|0 \right\rangle_{F_{\mu\nu}} = & & e_\psi \int_0^1 du F_{\mu\nu}(\xi x)
\left[d(u)+x^2d'(u)\right] \nm \\
&+& e_\psi\int_0^1 du \left(F_{\mu\rho}(\xi x) x^\rho x_\nu-
F_{\nu\rho}(\xi x)x^\rho x_\mu\right)b(u) \nm \\ \label{t2chiodd}
\eq
$x^2$ is not set to zero in front of $d'(u)$, because this wave function is a twist-4 correction to $d(u)$.

Matrix element $\left\langle 0 |\ov{\psi}(x)\psi(-x)|0 \right\rangle_{F}$ is zero in first order in the electric charge.

%%%%%%%%%%%%%%%%%%%%%%%%%%%%%%%%%%%%%%%%%%%%%%%%%%%%%

\subsection{Three-points wave functions}

Here, I build the classification in analogy to Ball \& Braun\cite{BBS}.

Chiral even:
\bq
\left\langle 0 | \ov{\psi}(x)g_s \Gtilde_{\mu\nu}(vx) \gamma_\alpha\gamma_5 \psi(-x)|0\right\rangle_{F_{\mu\nu}} &=& e_\psi
\int {\cal D} \unal F_{\mu\nu}([x\unal])q_\alpha A(\unal) \\
\left\langle 0 | \ov{\psi}(x)g_s G_{\mu\nu}(vx) i\gamma_\alpha\psi(-x)|0 \right\rangle_{F_{\mu\nu}} &=& e_\psi
\int {\cal D} \unal F_{\mu\nu}([x\unal])q_\alpha V(\unal)
\eq
where $\unal=\{\alpha_1,\alpha_2,\alpha_g\}$ is a set of momentum fractions.  The integration measure is defined as 
\be
\int {\cal D}\underline{\alpha} \equiv \int_0^1 d\alpha_1
  \int_0^1 d\alpha_2\int_0^1 d\alpha_g \,\delta\left(1-\sum \alpha_i\right)
\ee
and
\be
F_{\mu\nu}([x\unal])\equiv F_{\mu\nu}(x(\alpha_1-\alpha_2+v\alpha_g))
\ee

The different wave functions are classified by their different projections onto light-cone component, see Table~1 (the symbol $\perp$ means projection onto the plane perpendicular to $x$ and $q$). When one compares this list of distributions with the one for the vector meson (in \cite{BBS}), one can see that there is less possibilities in the case of the photon. It comes from the electromagnetic gauge invariance.

\begin{table}
\addtolength{\arraycolsep}{3pt}
\renewcommand{\arraystretch}{1.4}

$$
\begin{array}{|c|lcc|lc|}
\hline
{\rm Twist} &(\mu\nu\alpha) 
& \bar\psi \widetilde{G}_{\mu\nu}\gamma_\alpha\gamma_5
\psi & \bar\psi G_{\mu\nu}\gamma_\alpha \psi 
&(\mu\nu\alpha\beta) & \bar\psi G_{\mu\nu} \sigma_{\alpha\beta}\psi 
\\ \hline
3 & \cdot\perp \cdot &  A &  & \cdot\perp
   \cdot\perp & \\ \hline
4 & &  & & 
\perp\perp\cdot\!\perp & T_1
\\
 &   &  & & \cdot\perp
 \perp\perp & T_2 \\ 
& & & & \cdot * \cdot\perp & T_3\\
& & & & \cdot\perp\! \cdot\, * & T_4\\\hline
\end{array}
$$
\addtolength{\arraycolsep}{-3pt}
\renewcommand{\arraystretch}{1}
\caption[]{
Identification of three-particle wave functions with 
projections onto different light-cone components of the 
nonlocal operators. For example, $\cdot\perp\perp$ refers to
$\bar\psi \widetilde{G}_{\cdot\perp}\gamma_\perp\gamma_5\psi$.
}
\end{table}

Chiral odd:
\bq
\lefteqn{\left\langle 0 | \ov{\psi}(x)\sigma_{\alpha\beta} g_s G_{\mu\nu}(vx) \psi(-x)|0\right\rangle_{F_{\mu\nu}}=} \nm \\
& &e_\psi\int{\cal D}\unal \left\{\left[ F_{\mu\alpha}([x\unal])\gp_{\beta\nu}-
\frac{1}{qx}F_{\alpha\rho}([x\unal])x^\rho\gp_{\beta\mu}q_\nu - 
(\alpha \leftrightarrow \beta)\right]-(\mu \leftrightarrow \nu) \right\}
T_1(\unal) \nonumber \\
&+&e_\psi\int{\cal D}\unal \left\{\left[ F_{\mu\alpha}([x\unal])\gp_{\beta\nu}+
\frac{1}{qx}F_{\mu\rho}([x\unal])x^\rho\gp_{\alpha\nu}q_\beta - 
(\alpha \leftrightarrow \beta)\right]-(\mu \leftrightarrow \nu) \right\}
T_2(\unal) \nonumber \\
&+&e_\psi\int{\cal D}\unal F_{\alpha\beta}([x\unal])
\frac{1}{qx}(q_\mu x_\nu - q_\nu x_\mu)
T_3(\unal) \nonumber \\
&+&e_\psi\int{\cal D}\unal F_{\mu\nu}([x\unal])
\frac{1}{qx}(q_\alpha x_\beta - q_\beta x_\alpha)
T_4(\unal) \nonumber \\
\eq
\be
\left\langle 0 | \ov{\psi}(x)g_s G_{\mu\nu}(vx) \psi(-x)|0\right\rangle_{F_{\mu\nu}}=
e_\psi\int{\cal D}\unal F_{\mu\nu}([x\unal])S(\unal)
\ee
\be
\left\langle 0 | \ov{\psi}(x)ig_s \Gtilde_{\mu\nu}(vx) \psi(-x)|0\right\rangle_{F_{\mu\nu}}=
e_\psi\int{\cal D}\unal F_{\mu\nu}([x\unal])\widetilde{S}(\unal)
\ee

%%%%%%%%%%%%%%%%%%%%%%%%%%%%%%%%%%%%%%%%%%%%%%%%%%%%%%%%%%
%%%%%%%%%%%%%%%%%%%%%%%%%%%%%%%%%%%%%%%%%%%%%%%%%%%%%%%%%%

\sect{Some explicit expressions}

In this part, I use the results of \cite{BA, BBK} to obtain some evaluations of wave functions. The technic  is based on the conformal expansion\cite{BBK, BF, Ma, GS, Oh} (expansion on different parts which renormalized multiplicatively).

For the case of two-points chiral-odd wave functions, expressions can be extracted from \cite{BA}:

\bq
 \langle \bar \psi(0)\sigma_{\alpha\beta}
\psi(x)\rangle_F &=&
 e_\psi \langle\bar \psi\psi\rangle
\int_0^1 du\,F_{\alpha\beta}(ux)
\Big[\chi\phi_\gamma(u)+{x^2}g_\gamma^{(1)}(u)\Big]
\nonumber\\
&+&{}e_\psi \langle\bar \psi\psi\rangle
\int_0^1 du\,g_\gamma^{(2)}(u)\Big[x_\beta x_\eta F_{\alpha\eta}-
x_\alpha x_\eta F_{\beta\eta}-x^2 F_{\alpha\beta}\Big](ux) \nm \\
\eq
with
\begin{eqnarray}
  \phi_\gamma(u) &=& 6u(1-u) \nm \\
    g_\gamma^{(1)}(u) &=&-\frac{1}{8}(1-u)(3-u)
\nonumber\\
    g_\gamma^{(2)}(u) &=&-\frac{1}{4}(1-u)^2
\end{eqnarray}
and the value of $\chi$ (the magnetic susceptibility) is\cite{chi}:
\begin{equation}
  \chi(\mu= 1\,\text{GeV}) =-4.4 \,\text{GeV}^{-2}
\end{equation}

These relations can be transformed in order to obtain expressions for $d(u)$, $d'(u)$ and $b(u)$ (defined in equation \ref{t2chiodd}):
\bq
d(u)&=&\chi 6 u(1-u) \\
d'(u)&=&\frac{1}{8}u(u-2) \\
b(u)&=&-\frac{1}{4}u^2
\eq

For the case of three-points chiral-odd wave functions, the results of \cite{BBK} can be used:
\bq
\lefteqn{\left\langle 0|1/2\ov{\psi}(ux)\left[\sigma_{\alpha\xi}gG_{\beta\xi}
-\sigma_{\beta\xi}gG_{\alpha\xi}\right](vx)\psi(x)|0\right\rangle_
{F_{\mu\nu}}=} \nm \\
& &e_\psi \int_0^1 d\alpha_1 d\alpha_2 d\alpha_3 \delta(\Sigma \alpha_i -1)
\phi_\Sigma(\alpha_1\alpha_2\alpha_3)F_{\alpha\beta}(u\alpha_1 x+\alpha_2 x
+v\alpha_3 x) \\
\lefteqn{\left\langle 0|1/2\ov{\psi}(ux)\left[gG_{\alpha\beta}
-i\gamma_5g\Gtilde_{\alpha\xi}\right](vx)\psi(x)|0\right\rangle_
{F_{\mu\nu}}=} \nm \\
& &e_\psi \int_0^1 d\alpha_1 d\alpha_2 d\alpha_3 \delta(\Sigma \alpha_i -1)
\phi_\Gamma(\alpha_1\alpha_2\alpha_3)F_{\alpha\beta}(u\alpha_1 x+\alpha_2 x
+v\alpha_3 x)
\eq
with
\bq 
\phi_\Sigma(\alpha_i)&=&30(\alpha_1-\alpha_2)\left\{\alpha_3^2\left[
\kappa+\zeta_1(1-2\alpha_3)+\zeta_2(3-4\alpha_3)\right]+12\zeta_2
\alpha_1\alpha_2\alpha_3\right\}\left\langle\ov{\psi}\psi\right\rangle \\
\phi_\Gamma(\alpha_i)&=&30\alpha^2_3\left\{\kappa(1-\alpha_3)+\zeta_1
(1-\alpha_3)(1-2\alpha_3)+\zeta_2\left[3(\alpha_1-\alpha_2)^2-\alpha_3
(1-\alpha_3)\right]\right\}\left\langle\ov{\psi}\psi\right\rangle \nm \\
\eq
and
\bq
\kappa&=&0.2\\
\zeta_1&=&0.4\\
\zeta_2&=&0.3
\eq

These expressions imply:
\be
S(\unal)-\widetilde{S}(\unal)=15\alpha^2_3\left\{\kappa(1-\alpha_3)+\zeta_1
(1-\alpha_3)(1-2\alpha_3)+\zeta_2\left[3(\alpha_1-\alpha_2)^2-\alpha_3
(1-\alpha_3)\right]\right\}\left\langle\ov{\psi}\psi\right\rangle
\ee
\bq
\lefteqn{\left(T_1(\unal)+T_2(\unal)-\frac{1}{2}T_3(\unal)+\frac{1}{2}T_4(\unal)\right)=} \nm \\
& &30(\alpha_1-\alpha_2)\left\{\alpha_3^2\left[
\kappa+\zeta_1(1-2\alpha_3)+\zeta_2(3-4\alpha_3)\right]+12\zeta_2
\alpha_1\alpha_2\alpha_3\right\}\left\langle\ov{\psi}\psi\right\rangle
\nm \\
\eq

%%%%%%%%%%%%%%%%%%%%%%%%%%%%%%%%%%%%%%%%%%%%%
%%%%%%%%%%%%%%%%%%%%%%%%%%%%%%%%%%%%%%%%%%%%%

\sect{Conclusion}

In this paper, I classified the different wave functions of the photon and extracted some explicit expressions. An evaluation of all these wave functions (using the technic of conformal expansion) will be certainly very useful, in order to understand better the QCD at low energies and to use with a good accuracy the method of QCD light-cone sum rules for decay with photons.

\section*{Acknowledgments}

This work is supported by Schweizerischer Nationalfond. I am grateful to V. Braun for introducing me to the subject.

\end{document}